**Structured Driving-State Narratives for Small Language Model-Based GNSS Spoofing Detection**

**Abyad Enan***
Glenn Department of Civil Engineering
Clemson University, Clemson, South Carolina, 29634, United States
Email: aenan@clemson.edu

**Sagar Dasgupta, Ph.D.**
Research Engineer
Department of Civil, Construction, and Environmental Engineering
University of Alabama, Tuscaloosa, AL 35487, United States
Email: sdasgupta@ua.edu

**Mizanur Rahman, Ph.D.**
Assistant Professor
Department of Civil, Construction, and Environmental Engineering
University of Alabama, Tuscaloosa, AL 35487, United States
Email: mizan.rahman@ua.edu

**Mashrur Chowdhury, Ph.D.**
Eugene Douglas Mays Chair of Transportation
Glenn Department of Civil Engineering
Clemson University, Clemson, South Carolina, 29634, United States
Email: mac@clemson.edu

*Corresponding Author

## ABSTRACT

**Objectives:** Autonomous vehicles (AVs) depend on reliable Global Navigation Satellite System (GNSS) positioning. However, spoofed GNSS signals can induce plausible but incorrect vehicle states. This study develops a small language model (SLM)-based framework for detecting and classifying GNSS spoofing attacks by comparing vehicle behaviors independently derived from GNSS and other sensing sources.

**Methods**: The framework converts independent driving states from GNSS and other sensing sources into structured semantic narratives that are provided to an SLM for spoofing detection and attack classification. The performance of the SLM-based framework is compared with large language models (LLMs) fine-tuned on identical training data and evaluated on the same test set. The evaluation considers five classes: no attack, overshoot attack, stopped attack, turn-by-turn attack, and wrong-turn attack. The framework is also evaluated with geographically unseen field data collected in Clemson, South Carolina, United States.

**Findings**: Experimental results indicate that the evaluated SLMs achieve performance similar to the LLMs, achieving an average accuracy of 96.99%, precision of 99.05%, recall of 95.59%, and F1-score of 97.18%. In terms of computational efficiency and resource utilization, the SLMs demonstrate advantages over the LLMs by requiring lower inference latency and less GPU memory during both fine-tuning and inference. Evaluation using field data collected in a geographically distinct location further demonstrated its efficacy.

**Novelty**: This study develops an SLM-based GNSS spoofing detection framework that transforms independently derived GNSS and other onboard sensing source vehicle states into structured textual representations. This approach enables a compact language model to jointly learn the semantic and numerical inconsistency patterns associated with multiple sophisticated GNSS spoofing attacks.

**Practical Applications:** The presented framework can detect and classify GNSS spoofing attacks in real-time while requiring relatively low computational and memory resources, and is therefore suitable for deployment on resource-constrained vehicular computing platforms.

## INTRODUCTION

Global Navigation Satellite Systems (GNSS) play an essential role in providing precise position, velocity, and timing (PVT) solutions for transportation systems. Autonomous vehicles (AVs), a critical component of modern transportation systems, rely heavily on GNSS-derived PVT information for lane-level localization, route planning, trajectory following, turning decisions, and other safety-critical driving functions. Therefore, trustworthy PVT solutions are essential for the safe operation of AVs. However, GNSS is vulnerable to cyberattacks because civilian GNSS signals are unencrypted and openly accessible (Dunn and Disl 2012). Since low-cost software-defined radios are readily available, GNSS spoofing attacks are becoming easier to launch than ever before (K. (Curtis) Zeng et al. 2018). An attacker can transmit counterfeit GNSS signals to a receiver, causing it to estimate an incorrect location. Consequently, a spoofed AV receives a falsified PVT solution that can lead to incorrect localization, unnecessary route recalculation, inappropriate driving maneuvers, and erroneous turning or lane-changing decisions, ultimately compromising vehicle safety.

GNSS spoofing attacks can be broadly categorized as simplistic, intermediate, or sophisticated based on the attacker's knowledge of the target receiver's position and motion and the manner in which counterfeit signals are generated (Humphreys et al. 2008). Simplistic attacks generally transmit high-power, unsynchronized counterfeit signals and require little knowledge of the target's movement. Intermediate attacks use synchronized signals and may incorporate limited estimates of the receiver's motion. Both categories are comparatively easy to detect because they often produce abrupt changes in positioning, abnormal signal power, or signals arriving from a common direction. In contrast, sophisticated spoofing attacks require accurate knowledge of the target receiver's position and motion. The attacker transmits synchronized counterfeit signals using multiple coordinated transmitters or an antenna array, allowing the spoofed signals to closely reproduce the geometric distribution of authentic satellite signals. These attacks are therefore the stealthiest and most difficult to detect. If the attacker also knows the planned route, the manipulated trajectory can remain road-consistent and avoid abrupt discontinuities. Such trajectory-level effects are difficult to identify using checks that depend only on sudden changes in position or signal quality. This study, therefore, focuses on smooth navigation-domain manifestations of successful spoofing rather than on radio-frequency attack generation.

Existing GNSS spoofing detection approaches have made substantial progress. However, they still have several limitations when confronted with sophisticated attacks. Methods that focus on GNSS observables and signal quality are generally more effective at detecting abrupt anomalies than gradual carry-off attacks that preserve apparently plausible signal behavior (Kujur et al. 2024). Multi-sensor methods compare GNSS with IMU, odometry, LiDAR, or other sensor measurements, but coordinated attacks can exploit sensor-fusion logic while corrupting the estimated state, as demonstrated by (Shen et al. 2020). Vision-assisted driving behavioral validation can further strengthen positioning integrity; however, approaches based on visual-inertial odometry or camera-to-map matching may require considerable computation and prebuilt environmental information, while their reliability can be affected by changes in lighting, weather, and scene appearance (Elghazaly et al. 2023).

These limitations motivate a complementary approach that evaluates the semantic consistency of vehicle behavior rather than relying exclusively on raw signal characteristics, numerical residual thresholds, or computationally intensive localization pipelines. Nevertheless, it remains unclear whether heterogeneous GNSS and onboard sensor observations can be transformed into compact, structured driving narratives that enable a language model to distinguish legitimate vehicle behavior from sophisticated, spoofing attacks. In particular, it is unknown whether a single small language model (SLM) can jointly interpret structured descriptions of vehicle motion, maneuver, speed, and heading to distinguish legitimate driving from multiple spoofing attacks. Furthermore, the detection accuracy, end-to-end latency, robustness to errors in semantic state extraction, and the generalization of such a narrative-based framework to unseen routes and attack realizations remain insufficiently understood. This study addresses these gaps by developing an SLM-based GNSS spoofing detection framework that converts GNSS and other onboard sensor measurements into structured descriptions of vehicle motion, maneuvers, speed, and heading, and evaluates an SLM for low-latency detection and classification of sophisticated GNSS spoofing attacks.

SLMs are compact pretrained language models designed to provide language understanding capabilities with lower memory requirements and inference costs than large language models (LLMs). Recent studies have demonstrated the feasibility of deploying sub-billion- to few-billion-parameter language models on resource-constrained and mobile platforms, making them attractive for applications requiring local, low-latency processing (Abdin et al. 2024; Liu et al. 2024). Because they contain fewer parameters than LLMs, SLMs are more flexible for application-specific fine-tuning. Their lower memory requirements and inference costs also make them well-suited for edge-computing applications such as autonomous driving.

Motivated by these advantages, this study addresses the following research questions: (1) Can an SLM use structured driving-state narratives derived from GNSS and other onboard sensors to distinguish normal vehicle operation from sophisticated GNSS spoofing attacks? (2) How accurately can an SLM distinguish among stopped, overshoot, turn-by-turn, and wrong-turn spoofing attacks? (3) Can the proposed method provide sufficiently low detection latency for onboard vehicle monitoring? (4) How well does the developed method generalize to routes and attack realizations that were not observed during model development?

In this study, SLM does not process raw sensor time-series data. Instead, GNSS and onboard sensor measurements are converted into a structured representation of movement, maneuvers, speed and heading. This representation enables the SLM to evaluate the semantic consistency between independently derived vehicle states and produce a structured attack classification. The compact size and low inference latency of the SLM, therefore, make it a promising candidate for onboard GNSS integrity monitoring. The key contributions of this study are as follows:

- An SLM-based GNSS spoofing detection framework is developed by evaluating the consistency between vehicle states independently derived from GNSS and other onboard sensor measurements.
- A structured semantic representation preserves the source identity of movement, maneuver, speed, and heading fields, allowing one SLM to evaluate categorical and numerical inconsistencies without processing raw sensor time series.
- An independently collected real-world validation dataset obtained by driving an instrumented vehicle in Clemson, South Carolina, enables evaluation of the proposed framework on previously unseen routes and assessment of its generalization capability.

## RELATED STUDIES

Early GNSS spoofing defenses primarily examined signal characteristics within the receiver. Common detection indicators include abnormal received power, automatic gain control values, correlation-function distortion, unexpected auxiliary correlation peaks, carrier-phase inconsistencies, and common signal directions of arrival. Wesson et al. developed a power-distortion detector that jointly evaluates received power and correlation-function distortion to distinguish authentic signals, multipath, jamming, and carry-off spoofing (Wesson et al. 2018). Spatial processing provides another category of receiver-level defense. Because signals transmitted from geographically separated satellites arrive from different directions, whereas counterfeit signals are frequently emitted from a common antenna, antenna arrays and synthetic-aperture techniques can identify inconsistencies in signal geometry. Rothmaier et al. developed a signal-geometry-based framework that analyzes spatial characteristics to detect spoofed signals (Rothmaier et al. 2021). Although signal-level and spatial methods can provide strong protection, many require access to raw receiver measurements, modified tracking channels, multiple antennas, or specialized signal-processing capabilities. Furthermore, an advanced attacker using coordinated transmitting antennas, signal nulling, or carefully controlled power may reduce the effectiveness of an individual detection indicator. The simultaneous use of complementary defenses has therefore been recommended for protection against sophisticated attacks (Psiaki and Humphreys 2016).

An alternative approach is to validate the GNSS solution using measurements from sensors that are independent of the received satellite signals. Inertial sensors are particularly useful because an external GNSS spoofer cannot directly manipulate their measurements. Swaszek et al. used shipboard IMU measurements and generalized likelihood ratio testing to compare GNSS-derived motion with vessel dynamics (waszek et al. 2014). More recent studies have combined inertial information with precise GNSS observables. Clements et al. proposed a carrier-phase and IMU-based detector for ground vehicles that monitors the residual cost of a tightly coupled carrier-phase differential GNSS estimator (Clements et al. 2022). The method exploits the practical difficulty faced by an attacker in predicting small vehicle motions caused by roadway irregularities and detected injected worst-case attacks.

Machine-learning methods have increasingly been applied to GNSS spoofing detection because they can learn multidimensional attack patterns without requiring a manually specified threshold for every measurement. Semanjski et al. used multiple GNSS observables as inputs to supervised machine-learning models for detecting spoofing and meaconing (Semanjski et al. 2020). Deep-learning methods have also been developed for signal-level detection. Borhani-Darian et al. proposed a deep-learning classifier combined with a clustering procedure to detect spoofing and estimate the number and characteristics of counterfeit signals (Borhani-Darian et al. 2024). Other studies have used neural networks to learn differences between GNSS- and IMU-derived velocity estimates. Guizzaro et al., for example, constructed feature vectors from GNSS and inertial velocity differences and trained a neural network to identify manipulated trajectories (Guizzaro et al. 2022). Although these methods reduce dependence on manually developed detection rules, they generally operate on receiver observables or numerical feature vectors and provide limited semantic interpretation of the combined vehicle behavior.

For autonomous-vehicle applications, Dasgupta et al. developed a sensor-fusion framework that uses an LSTM network to predict vehicle location shifts from in-vehicle sensors and compares those predictions with GNSS-derived displacement (Dasgupta et al. 2022). The framework also uses motion-state recognition and separate turn-detection methods to identify turn-by-turn, overshoot, stopped, and wrong-turn attacks. The approach demonstrated that sophisticated trajectory-level attacks can be identified by comparing GNSS behavior with independently observed vehicle dynamics. A subsequent field evaluation used GNSS and IMU data collected from an instrumented vehicle and confirmed the feasibility of detecting turn-by-turn, wrong-turn, and slow-drift manipulations in a real road environment (Dasgupta et al. 2024). However, coordinated attacks can exploit synchronization and sensor-fusion logic to maintain internal consistency while corrupting the estimated vehicle state, as demonstrated by (Shen et al. 2020).

Recent advances in foundation models have enabled heterogeneous sensor data to be represented within a shared semantic space. Aldeen et al. developed a vision-language model framework that combines front-camera images, in-vehicle sensor measurements, and GNSS-derived maneuvers to detect wrong-turn, stopped, and overshoot attacks (Aldeen et al. 2026). The framework identifies inconsistencies in vehicle motion and maneuver states, such as moving versus stationary and turning versus traveling straight. However, it cannot detect more complex attacks, such as turn-by-turn spoofing, which requires evaluating speed inconsistencies between GNSS and independent sensors, since both . However, it cannot detect more complex attacks, such as turn-by-turn spoofing, because both the IMU and the compromised GNSS indicate that the vehicle is moving; detection therefore requires evaluating discrepancies in their estimated speeds. The framework may also fail to distinguish a spoofed U-turn from a left turn because both initially involve leftward motion, while the method does not account for the magnitude of the heading change. In addition, the framework has relatively high computational requirements.

The reviewed literature demonstrates substantial progress in receiver-level monitoring, GNSS/INS consistency checking, sensor fusion, and data-driven spoofing detection. Nevertheless, several gaps remain. Signal-level techniques often require access to specialized GNSS observables or receiver modifications, while multi-sensor frameworks may employ separate algorithms and manually defined rules for motion, displacement, and maneuver inconsistencies. Existing machine- and deep-learning approaches primarily process raw numerical features and have not extensively investigated whether compact language models can jointly interpret categorical driving states and their associated numerical measurements. Moreover, computational efficiency are not consistently evaluated across models of different sizes.

This study addresses these gaps by independently deriving vehicle behavior from GNSS and other sensor measurements and converting the resulting movement, maneuver, speed, and heading states into a structured textual narrative. A fine-tuned SLM then performs unified multiclass classification of normal behavior and four sophisticated spoofing scenarios. By comparing SLMs with larger language models using identical training and testing data, evaluating computational latency and memory consumption, and testing the selected model on geographically unseen field data, this study examines both the detection capability and practical deployment potential of language-model-based GNSS spoofing detection.

## ATTACK SCENARIOS AND ASSUMPTIONS

The framework is evaluated using four trajectory-level attack scenarios that emulate the navigation outputs produced after successful spoofing. The scenarios are stopped, overshoot, turn-by-turn, and wrong-turn attacks. In each case, the GNSS-reported position is manipulated while the vehicle's physical motion and onboard sensor measurements remain unchanged. The attacker is assumed to know the target vehicle's route and approximate motion. The attack scenarios are presented in **Figure 1**.

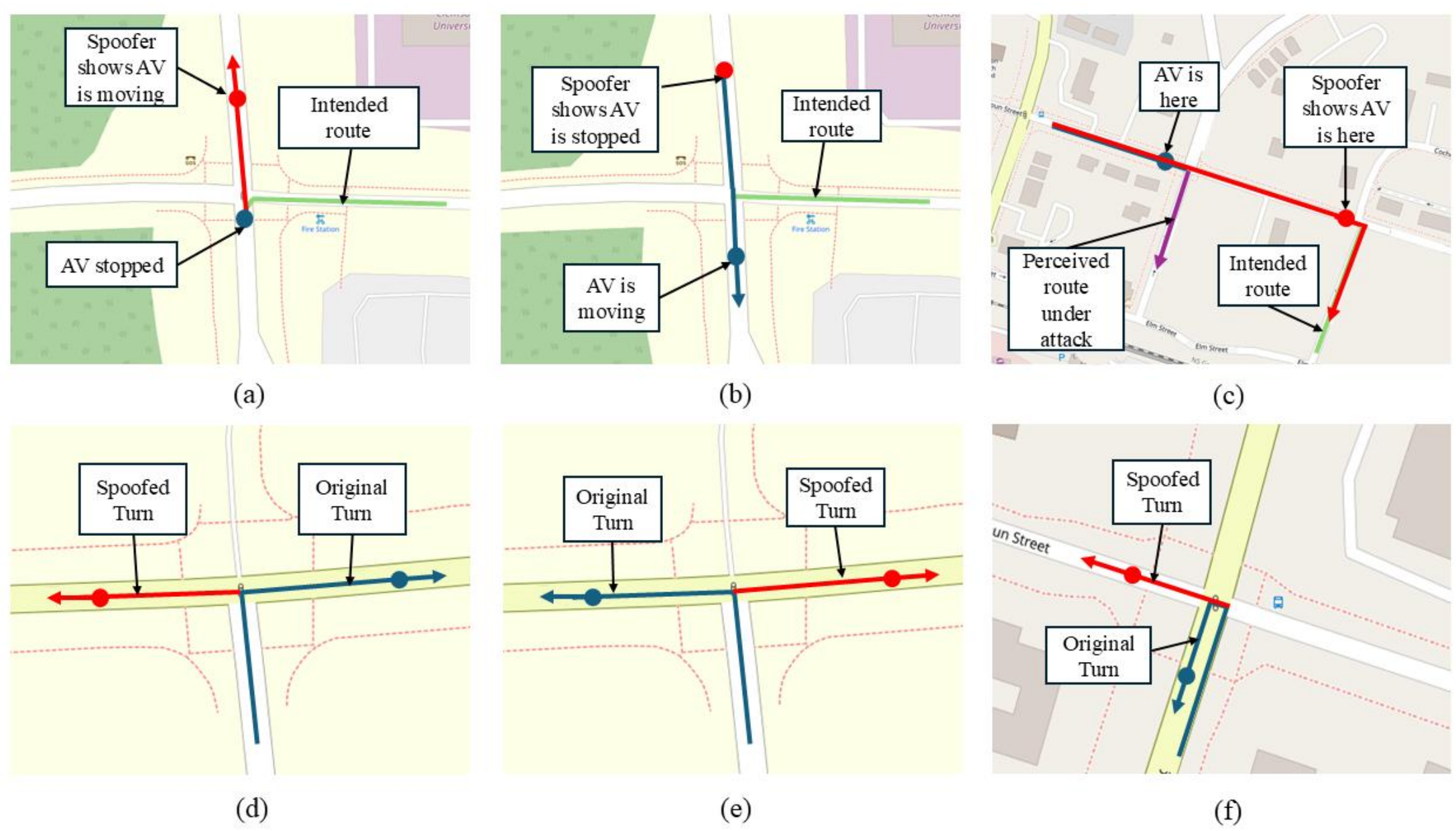

**Figure 1 Attack models considered in this study: (a) stopped, (b) overshoot, (c) turn-by-turn, (d) wrong turn (right turn), (e) wrong turn (left turn), and (f) wrong turn (U-turn) attack**

Stopped attack: A stopped attack is initiated while the vehicle is physically stationary, such as when it is waiting at a stop sign, a traffic signal, or in a traffic queue. During the attack, the adversary generates a sequence of counterfeit GNSS positions that appear to advance along the roadway (Merwe et al. 2018). The navigation system, therefore, interprets the vehicle as moving even though the speedometer and inertial sensors indicate no physical motion. This false movement may cause the navigation application to advance prematurely along the planned route, incorrectly determine that an intersection has been crossed, or provide instructions that are inconsistent with the vehicle's actual surroundings.

Overshoot attack: An overshoot attack creates the opposite inconsistency of a stopped attack. The vehicle continues moving while the compromised GNSS receiver repeatedly reports the same or nearly the same position (Merwe et al. 2018). Consequently, the navigation system perceives the vehicle as stationary or positioned behind its actual location, which may cause it to issue a maneuver instruction too late, after the vehicle has already passed the relevant intersection or exit.

Turn-by-turn attack: A turn-by-turn attack assumes that the attacker possesses information about the vehicle's planned route, destination, or sequence of navigation maneuvers (K. C. Zeng et al. 2017). Instead of introducing an arbitrary position error, the attacker constructs a plausible false trajectory corresponding to a different set of roads. The spoofed route is designed to preserve characteristics of the intended journey, such as the number, direction, timing, or spacing of turns. As a result, the manipulated

GNSS and onboard sensor streams can therefore indicate the same general maneuver but report inconsistent speed and route progression, causing turns to appear earlier or later than they occur physically.

Wrong turn attack: Near an intersection, the spoofed GNSS trajectory represents a maneuver different from the vehicle's physical turn. An actual right turn may be represented as a left turn, or an actual U-turn may be represented as a conventional left turn. These scenarios can preserve plausible speed and displacement while producing inconsistent maneuver direction or heading change.

**System Assumptions and Attacker Capabilities**

This study considers an adversary capable of producing gradual, road-consistent changes in the position solution reported by the vehicle's GNSS receiver. The attacker is assumed to know the planned route, expected maneuvers, and approximate motion. The study neither synthesizes nor transmits radio-frequency signals and does not evaluate receiver tracking-loop behavior. Only the GNSS-reported position sequence is modified during attack generation like previous studies by (Dasgupta et al. 2022). The original timestamps and onboard sensor measurements are preserved and treated as trustworthy. Speed, heading, movement, and maneuver states for the manipulated GNSS stream are then derived from the modified coordinates.

**Attack Data Generation**

A successful spoofer can manipulate the navigation solution produced by a GNSS receiver. This study emulates the trajectory-level effects of such manipulation by modifying GNSS-derived coordinates. Detection is based on inconsistencies in movement, maneuver direction, speed, heading, and progression along the road network rather than on abnormalities in the received waveform. Radio-frequency signal synthesis and receiver-level tracking analysis are outside the scope of the study. Attack scenarios are generated by modifying the GNSS coordinates while preserving timestamps and the measurements from the onboard sensor stream. The generated positions are matched to the surrounding OpenStreetMap road network so that the manipulated trajectories follow valid road segments.

For a stopped attack, stationary roadway segments are identified, and a sequence of synthetic GNSS coordinates is generated ahead of the physical stopping location. The coordinates progress gradually along the selected road segment, as shown in **Figure 1(a)**. The GNSS stream, therefore, indicates movement while the onboard sensor stream remains stationary. For an overshoot attack, a moving trajectory is selected, and the GNSS-reported position is held constant near an intersection or stop location. The GNSS stream, therefore, indicates a stationary vehicle while the onboard sensor stream indicates continued forward motion, as shown in **Figure 1(b)**. Wrong-turn attacks are generated from genuine left turns, right turns, and U-turns. At the start of the maneuver, the original GNSS trajectory is replaced by a road-valid trajectory representing a different turn. The post-intersection travel distance is kept close to that of the original trajectory to avoid unrealistic implied speeds, as illustrated in **Figure 1(d)** through **Figure 1(f)**. The turn-by-turn attack is generated using premature-progress and delayed-progress cases. In the premature case, the spoofed coordinates advance faster than the physical vehicle. In the delayed case, they progress more slowly. Coordinate spacing is adjusted to create a discrepancy between GNSS-derived and onboard-sensor-derived speed while preserving the original timestamps and road-consistent headings.

# DATA PREPARATION

**KITTI Benchmark Suite**

The KITTI dataset is used for model development and held-out evaluation (Geiger et al. 2013). It contains synchronized driving data collected in Karlsruhe, Germany, using a vehicle equipped with cameras, LiDAR, and an OXTS RT3003 GNSS and inertial navigation system. The dataset includes urban, rural, and highway driving conditions and provides the position and motion variables needed to construct the evaluated scenarios.

This study uses latitude and longitude to derive the GNSS position stream and uses the available OXTS motion-state outputs for the comparison stream. The recorded variables include position, heading, speed, acceleration, angular rate, accuracy indicators, satellite information, and navigation-status fields. The OXTS unit operates natively at 100 Hz, while the KITTI release provides measurements synchronized at 10 Hz (Geiger et al. 2013). The routes are plotted and segmented based on vehicle movement and maneuvers. Stopped, overshoot, turn-by-turn, and wrong-turn scenarios are then generated using the procedures described above. Ten attack scenarios from each attack category are randomly selected for held-out testing. The remaining attacked and non-attacked scenarios are used for fine-tuning and validation.

### Field Data

To assess the geographical generalizability of our framework, field data were collected using an instrumented vehicle in Clemson, South Carolina, United States. These data were collected exclusively for independent testing and were not used for model fine-tuning. The objective was to determine whether a framework fine-tuned on KITTI data collected in Germany could generalize to driving data from a different geographic region and country.

The field vehicle was instrumented with a Cohda Wireless MK6C data-acquisition platform, which included an integrated multi-constellation GNSS receiver and dual Controller Area Network (CAN) interfaces. The GNSS receiver provided latitude and longitude measurements, while the CAN interfaces provided vehicle-speed and heading-related kinematic measurements from the vehicle network. The GNSS and CAN data streams were recorded at 10 Hz to match the sampling frequency of the KITTI data used for model development. The vehicle was operated during two data-collection sessions: one during peak traffic conditions and the other during off-peak conditions. The collected data were segmented into maneuver categories: turning and straight driving, and motion-state categories: stopped and moving. Twenty driving scenarios were collected for each category. The four spoofing attacks described in the previous section: stopped, overshoot, turn-by-turn, and wrong-turn attacks, were subsequently generated using these field data. The complete field dataset, including attacked and non-attacked scenarios, was reserved exclusively for evaluating the fine-tuned framework.

## SLM-BASED ATTACK DETECTION FRAMEWORK

### System Overview

The framework detects GNSS spoofing attacks by examining the consistency between vehicle states independently obtained from GNSS and other GNSS-independent onboard sensors. The framework uses latitude and longitude coordinates reported by the GNSS receiver, together with vehicle speed and heading measurements obtained from other onboard sensors. For model development and evaluation of this study on the KITTI dataset, vehicle states derived from GNSS coordinates are compared with the corresponding motion states obtained from OXTS navigation measurements, which serve as an independent onboard motion-sensing source. The system architecture consists of four major components: (1) data acquisition and preprocessing, (2) driving state extraction, (3) narrative construction, and (4) SLM-based attack detection module. The framework is illustrated in **Figure 2**.

First, GNSS and onboard sensor measurements are collected during vehicle operation. The GNSS receiver provides a sequence of latitude and longitude coordinates, whereas the onboard sensors provides vehicle speed and heading information. The displacement between consecutive latitude and longitude observations is calculated to estimate the GNSS-based speed. Similarly, the direction of movement between consecutive coordinates is used to determine the GNSS-based heading. This produces two independently derived descriptions of vehicle behavior: one based on GNSS measurements and another based on onboard sensor observations.

The extracted measurements are organized into fixed-length temporal windows. Using a sequence of observations rather than a single measurement enables the framework to capture changes in vehicle movement and direction over time. For each window, the framework determines motion-related and maneuver-related characteristics, including whether the vehicle is moving or stopped, whether it is traveling straight or turning, its estimated speed, and its headings. These characteristics are calculated separately for the GNSS and other sensor data streams.

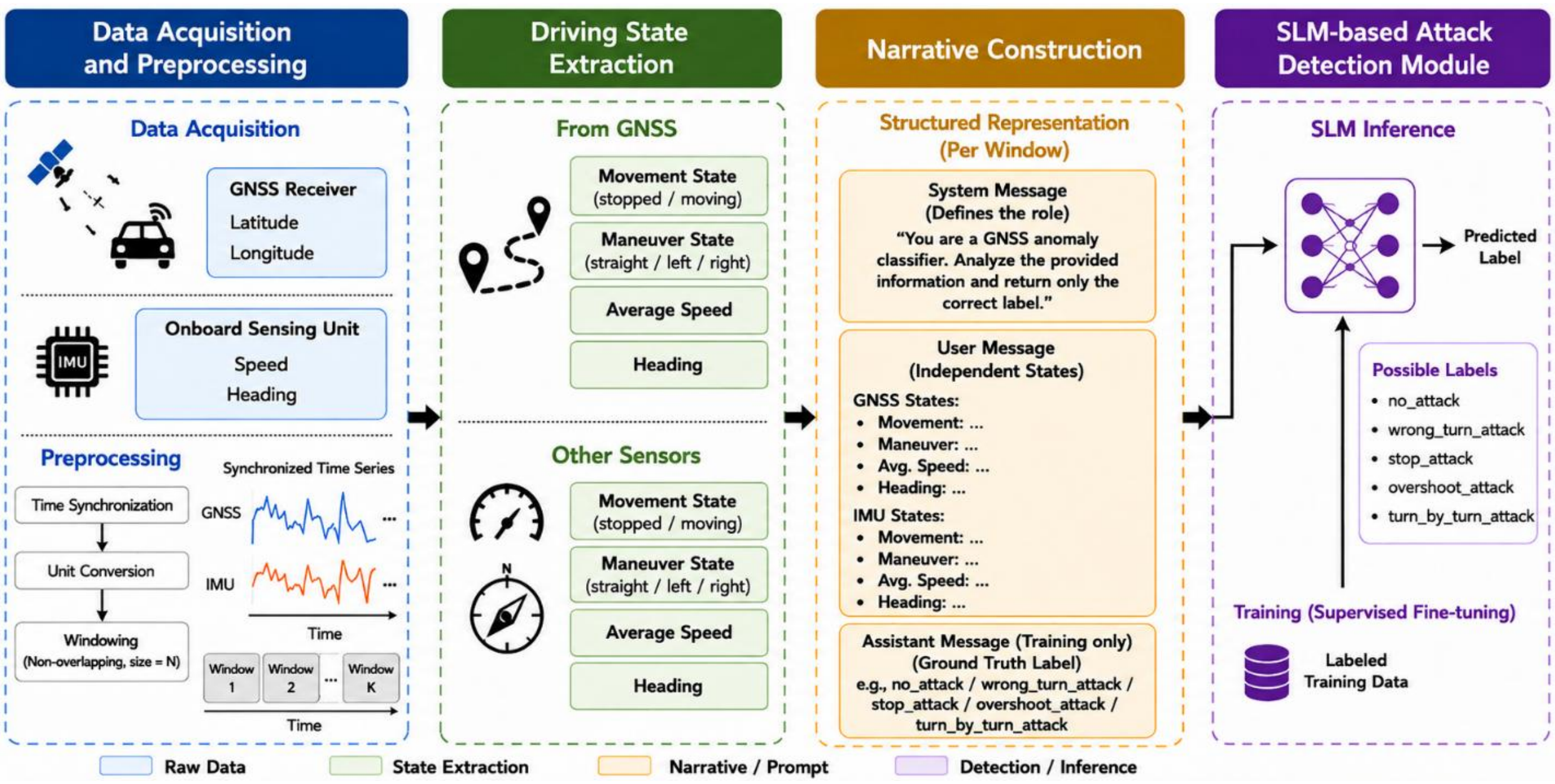

**Figure 2 GNSS spoof detection framework developed in this study**

The resulting features are converted into a structured representation that can be processed by an SLM. Each input instance describes the vehicle state inferred from GNSS, and the corresponding state obtained from the other sensors. The SLM evaluates the semantic consistency between the two independently derived vehicle states. For example, GNSS may indicate that the vehicle is moving while the motion sensor indicates that it is stopped, or the GNSS trajectory may represent a left turn while the sensor measurements indicate a right turn. Such disagreements may represent the behavioral effects of a spoofing attack. Based on the relationships among movement, maneuver, speed, and heading, the SLM assigns an attack label to each input window. The output may indicate normal vehicle operation or one of the spoofing scenarios considered, including stopped, overshoot, wrong-turn, and turn-by-turn attacks. In this manner, the architecture uses the learned classification capability of an SLM to identify inconsistencies between GNSS-derived navigation information and the vehicle motion measured by the other sensors.

## Data Acquisition and Preprocessing

The GNSS and other sensor measurements need to be synchronized to obtain two independent descriptions of vehicle motion if different sensors have different sampling frequencies. Since our data is already synchronized at 10 Hz, no further data synchronization is required. Each observation is associated with a timestamp to preserve the temporal relationship between the GNSS and sensor measurements. In this study, latitude and longitude are used as the GNSS inputs, and GNSS-based speed and heading are derived from consecutive coordinates using the following equations:

$$d = 2R\,\arcsin\left(\sqrt{sin^2\left(\frac{\Delta\phi}{2}\right) + \cos(\phi_1)\cos(\phi_2)sin^2\left(\frac{\Delta\lambda}{2}\right)}\right) \tag{1}$$

$$v_{GNSS} = \frac{d}{\Delta t} \tag{2}$$

$$\theta_{GNSS-radian} = atan2(\sin(\Delta\lambda)\cos(\phi_2), \cos(\phi_1)\sin(\phi_2) - \sin(\phi_1)\cos(\phi_2)\cos(\Delta\lambda)) \tag{3}$$

$$\theta_{GNSS} = \left(\frac{180}{\pi}\theta + 360\right) mod\ 360 \tag{4}$$

**Equation 1** is the Haversine formula (Robusto 1957) for distance ($d$) measurement between two successive coordinates ($\phi_1$, $\lambda_1$) and ($\phi_2$, $\lambda_2$) where latitude ($\phi$)and longitude ($\lambda$)are expressed in radians. R = 6,371 kilometers (km) is the Earth's radius. The vehicle speed $v_{GNSS}$ is computed by **Equation 2**, where $\Delta t = t_i - t_{i-1} = 0.1$ second (1/(10 Hz)) is the time taken to travel the distance $d$. The speed in km/s is then converted to mph by multiplying by 2236.94. **Equation 3** calculates the heading in radians for two successive coordinates, which is then converted to degrees using **Equation 4**.

**Driving State Extraction**

After data processing, the measurements are divided into fixed-length temporal windows. Each window contains consecutive timestamp data and represents a short interval of vehicle operation. Window-based processing reduces the effect of isolated sensor noise and enables continuous numerical measurements converted into meaningful descriptions of vehicle behavior that can be interpreted by the SLM.

The movement state is classified as either moving or stopped. The average speed within each window is calculated separately for GNSS and motion sensor. If the average speed is below a predefined threshold, the corresponding vehicle state is labeled as stopped. Otherwise, it is labeled as moving. We choose 1 mph as the threshold, since both GNSS and OXTS speed agree that the vehicle is stopped when the speed is below 1 mph in approximately 99% of unspoofed windows from the training set. Using the average speed over a temporal window reduces the influence of small GNSS position fluctuations and sensor noise, particularly when the vehicle is physically stationary.

The maneuver state is determined from the change in heading over the observation window. The heading at the beginning of the window is compared with the heading at the end of the window to estimate the direction and magnitude of the vehicle's rotation. When the heading change remains within a predefined tolerance, the vehicle is classified as traveling straight. A heading change exceeding the straight-driving tolerance is classified as either a left turn or a right turn according to the direction of rotation.

In this study, we determine the threshold using the ground-truth maneuver state from the training data. The ground-truth maneuvers are established by map-matching the GNSS coordinates to roadway links obtained from OSM and monitoring the corresponding road-link identifiers as the vehicle moves. A change between consecutive road-link identifiers indicates that the vehicle has transitioned from one roadway segment to another and is therefore considered a potential turning event, following the general navigation-based approach described by (Dasgupta et al. 2022; Zhao et al. 2018). The turn direction is determined using GNSS samples immediately before and after the detected link transition.

For each window, the heading change is calculated from the first and last heading measurements as follows: $\Delta\theta = \theta_{10} - \theta_1$ ; where $\theta_1$ is the heading at the beginning of the window and $\theta_{10}$ is the heading at the end of the window. The heading difference is adjusted to account for the circular nature of heading measurements, particularly when the heading crosses the 0°/360° boundary. For a given heading threshold $T$, a window is classified as a turning window when $|\Delta\theta| \geq T$. A turning window is subsequently classified as a right turn when the signed heading change is positive and as a left turn when the signed heading change is negative. In contrast, a window is classified as straight when: $|\Delta\theta| < T$. To identify an appropriate threshold, the percentage of turning and straight-driving windows satisfying these conditions was evaluated at different threshold values. For the known turning data, the agreement percentage is calculated as:

$$Turn\ agreement\ (\%) = \frac{\#turning\ windows\ satisfying\ |\Delta\theta| \geq T}{\#turning\ windows} \times 100 \tag{5}$$

For the known straight-driving data, the agreement percentage is calculated as:

$$Straight\ agreement\ (\%) = \frac{\#straight\ driving\ windows\ satisfying\ |\Delta\theta| \leq T}{\#straight\ windows} \times 100 \tag{6}$$

**Equation (5)** represents the percentage of actual turning windows that are correctly retained as turns when greater than the threshold. **Equation (6)** represents the percentage of actual straight-driving windows whose heading changes remain within the specified threshold.

To obtain the threshold, we measure the turn agreement percentages from **Equation (5)** and **Equation (6)** for different threshold values presented in **Figure 3**. **Figure 3** illustrates the relationship between the heading threshold and the agreement percentages for both maneuver categories, generated only from unspoofed training data. The turning agreement decreases as the threshold increases, whereas the straight-driving agreement increases. At a threshold of 10°, ≈95% of the turning windows exhibit an absolute heading change of at least 10°, while ≈95% of the straight-driving windows exhibit an absolute heading change below 10°. Therefore, 10° is selected as the heading-change threshold because it provides a balanced separation between turning and straight-driving maneuvers.

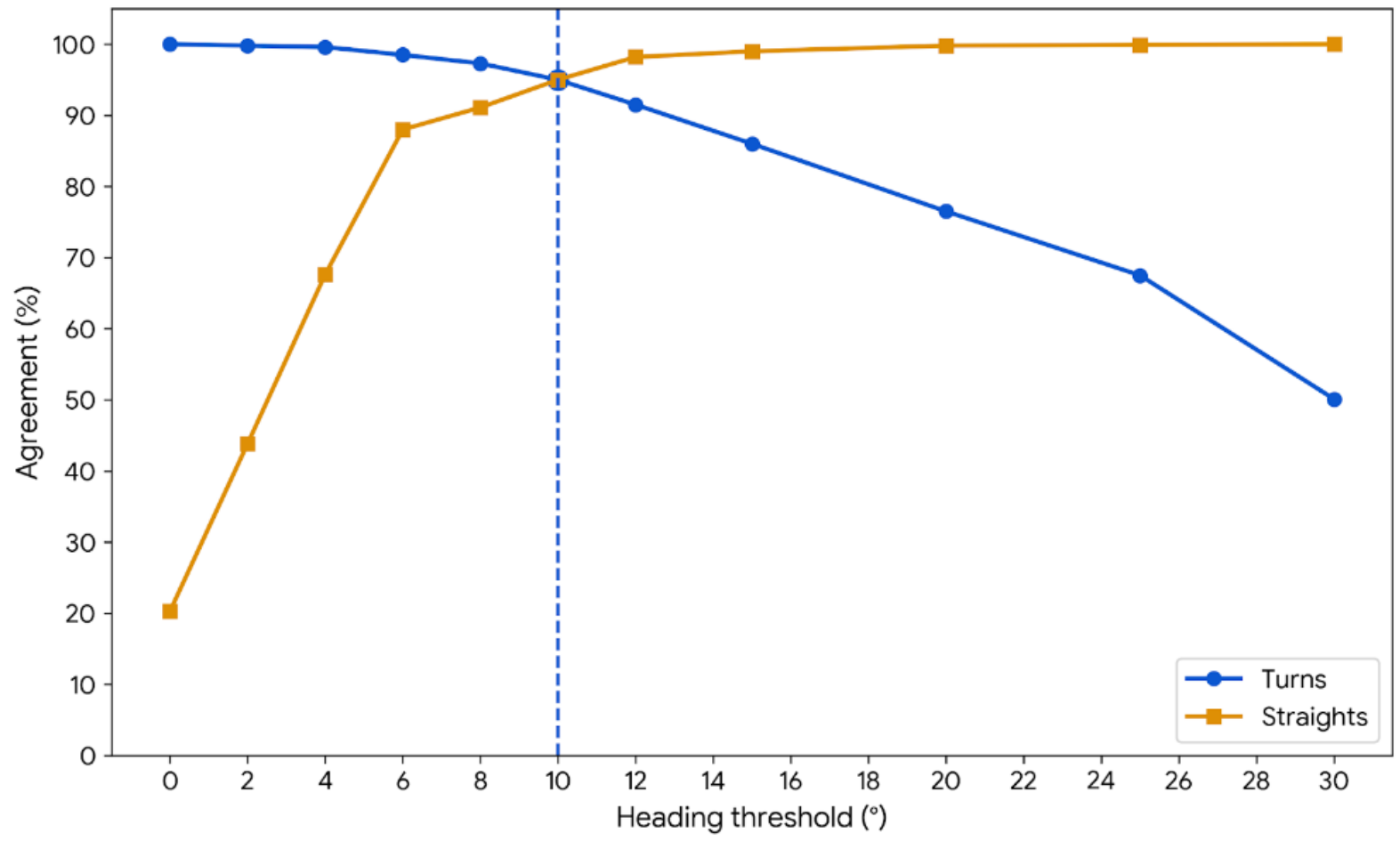

**Figure 3 Percentage of turning and straight-driving windows satisfying different absolute heading-change thresholds for an observation window of 10 samples**

**Narrative Construction**

The narrative reconstruction module converts the numerical and categorical driving states extracted from each observation window into a structured textual representation suitable for processing by the SLM. Instead of providing raw GNSS and sensors time-series measurements to the SLM, the framework represents each observation window as a structured set of textual key-value fields describing movement, maneuver, average speed over the window, and heading at the window's final timestamp. This process aggregates the time-series measurements within each observation window into a compact semantic representation, reducing input complexity while preserving the relationships relevant to spoofing detection. Narratives are constructed independently for the GNSS and sensor streams. Maintaining separate narratives prevents information from one sensing source from influencing the interpretation of the other before semantic consistency is evaluated.

A deterministic template is used to preserve a fixed field order, consistent attribute names, standardized categorical terms, and uniform numerical formatting across all observation windows. The general narrative structure can be represented as follows:

```
{
 "messages": [
   {
    "role": "system",
    "content": "You are a GNSS spoofing attack classifier. Analyze the provided GNSS-derived and
sensor-derived vehicle-state information and return only the correct class label."
   },
   {
    "role": "user",
```

"*content*": "*movement_GNSS: moving\nturn_GNSS: straight\nspeed_GNSS: 26.06 mph\nfinal_heading_GNSS: 344.01°\nmovement_sensor: moving\nturn_sensor: straight\nspeed_sensor: 26.13 mph\nfinal_heading_sensor: 343.69°*"
},
{
"*role*": "*assistant*",
"*content*": "*no_attac*k"
}
]
}

The assistant message containing “content” is not part of the constructed driving narrative. It is the ground-truth output used during supervised fine-tuning. During inference, only the system and user messages are supplied, and the model generates the assistant response. The narrative has these components:

The use of controlled templates provides a consistent vocabulary for movement and maneuver states, including moving, stopped, straight, left turn, and right turn. Numerical values are retained in the narrative to allow the SLM to examine not only categorical agreement but also the magnitude of differences between the two sensing modalities. For example, two narratives may both describe the vehicle as moving straight while reporting substantially different speeds. Similarly, the GNSS and sensor narratives may indicate opposite turning directions or considerably different headings.

**SLM-based Attack Detection Module**

The SLM-based attack detection module is the decision-making component of the proposed framework. It receives the structured representation generated for each observation window and determines whether the GNSS-derived vehicle state is consistent with the corresponding state observed by the other onboard sensors. The module performs multiclass classification, allowing it to distinguish normal driving observations from the different GNSS spoofing attack scenarios considered in this study. Each observation window is formatted using a chat-based structure consisting of system, user, and assistant messages. The system message defines the role of the SLM and instructs it to operate as a GNSS spoofing attack classifier. The user message contains the vehicle states independently derived from the GNSS and sensor data streams, including the movement state, maneuver state, average speed, and heading information. During training, the assistant message contains the ground-truth class associated with the observation window.

The input-output relationship for the detection module can be conceptually represented as:

$$Predicted\ label\ =\ SLM(system\ instruction, GNSS\ states, sensor\ states)$$

The user input contains the following structured attributes:
GNSS-derived information: *movement_GNSS, turn_GNSS, speed_GNSS, and final_heading_GNSS*
sensor-derived information: *movement_sensor, turn_sensor, speed_sensor, and final_heading_sensor*

The model is instructed to generate one of the predefined labels. The generated response is subsequently normalized and matched against the valid label set. The predefined label set represents the normal condition and the four spoofing attack scenarios considered in this study. Because the output is restricted to a predefined label, the module performs classification rather than generating an unrestricted textual explanation. This output constraint also simplifies the evaluation and prevents unnecessary variation in model responses.

During training, the SLM is presented with labeled examples of normal and spoofed vehicle behavior. Each example contains GNSS- and sensor-derived states for the same observation window, along with the corresponding ground-truth label. Through these examples, the model learns the expected semantic and numerical relationships between the two sensing modalities under normal operating conditions and under different spoofing attacks. During inference, only the system and user messages containing the GNSS- and sensor-derived states are provided to the fine-tuned SLM. The ground-truth assistant message is omitted, and the model generates the predicted label as its response. The generated output is then normalized and compared with the predefined class labels to obtain the final classification decision for the observation window.

For a normal observation, the GNSS and sensor representations are expected to exhibit compatible vehicle behavior. For example, both sources may indicate that the vehicle is moving straight with similar speeds and headings. The numerical values do not strictly need to be identical because GNSS and onboard sensor measurements may contain noise and estimation differences. The model must therefore learn to tolerate small discrepancies while preserving the no_attack classification. In contrast, spoofing attacks introduce characteristic inconsistencies between the two sources. A stopped attack may cause one source to indicate vehicle movement while the other indicates that the vehicle is stationary. A wrong-turn attack may produce conflicting maneuver states, opposite turning directions, or substantially different heading values. A turn-by-turn attack may preserve the general maneuver state while introducing an abnormal difference in speed, indicating that the spoofed GNSS trajectory is advancing earlier or later than the actual vehicle. An overshoot attack may cause the spoofed GNSS-derived state to indicate that the vehicle has slowed or stopped, while the sensor-derived state indicates continued motion. SLMs use the interrelationships among movement, maneuver, speed, and heading to distinguish these attack patterns.

The model is not explicitly provided with a separate rule for every possible disagreement. Instead, it learns the relationships between the structured input patterns and their corresponding class labels from representative training examples. For instance, a small difference between speed_GNSS and speed_sensor may occur during normal operation, whereas a substantially larger speed difference may indicate that the spoofed GNSS trajectory is progressing at a rate inconsistent with the actual vehicle motion. Similarly, minor heading differences may result from sensor uncertainty, while opposing maneuver directions or large heading deviations may indicate a wrong-turn attack. By considering all attributes collectively, the SLM evaluates the combined window-level driving context rather than making a decision based on an individual measurement.

## EXPERIMENTS AND RESULTS

### SLM Fine-Tuning

Two instruction-tuned SLMs are considered for the GNSS spoofing attack detection framework: Qwen3-1.7B and Llama-3.2-1B-Instruct. The models are selected because their relatively small parameter sizes (1.7 billion and 1.23 billion, respectively) make them suitable for resource-constrained and potentially edge-based deployment while retaining the instruction-following capabilities required to process structured textual inputs. The training data represents a supervised instruction-fine-tuning task in which each model learns to map a structured description of the vehicle state to one of the predefined classes. During fine-tuning, the ground-truth assistant response is included in each training sample. During validation and testing, only the system and user messages are provided, and the label generated by the model is treated as the predicted class. Before fine-tuning, each example is processed using the tokenizer and chat template associated with the corresponding base model. Because the models use model-specific tokenization and conversation formatting, the same semantic input is separately converted using each model's native chat template. This ensures that the system, user, and assistant roles are represented in the format expected by each model. Parameter-efficient fine-tuning is performed using Low-Rank Adaptation (LoRA). Rather than updating all parameters of the pretrained models, LoRA introduces a limited number of trainable low-rank parameters into selected model layers while keeping the original parameters frozen. This substantially reduces the computational and memory requirements of fine-tuning while preserving the knowledge acquired during pretraining, making it suitable for adapting compact language models to a specialized classification task using a limited domain-specific dataset.

### Baseline LLM Fine-Tuning

To assess whether larger model size provides a meaningful performance advantage, Qwen3-8B and Mistral-7B-Instruct-v0.3 are included as baseline LLMs. All models use the same data partitions, feature order, class definitions, output requirements, and LoRA-based fine-tuning procedure, while model-specific chat templates accommodate differences in tokenizer and architecture. This controlled comparison evaluates whether the larger models improve detection performance sufficiently to justify their greater training and inference costs.

**Evaluation of the Detection Framework**

**Tables 1** and **2** summarize the performance of the fine-tuned SLMs and LLMs on the held-out test set, where bold values indicate the best result in each column. Overall, all four models achieved high detection performance, with accuracy ranging from 96.42% to 96.99% and macro F1-scores ranging from 95.87% to 97.18%. Qwen3-1.7B achieved the highest macro precision of 99.05% and the highest macro F1-score of 97.18%, while sharing the highest accuracy of 96.99% with Mistral-7B. Mistral-7B obtained the highest macro recall of 95.89%. These results demonstrate that the smaller Qwen3-1.7B model performs comparably to, and in several metrics slightly better than the LLMs.

**TABLE 1 Overall performance of the evaluated SLMs and LLMs on the held-out test set**

| Group | Model | Accuracy (%) | Macro Precision (%) | Macro Recall (%) | Macro F1-score (%) |
|---|---|---|---|---|---|
| SLMs | **Qwen3-1.7B** | **96.99** | **99.05** | 95.59 | **97.18** |
| | **Llama-3.2-1B** | 96.42 | 97.67 | 94.54 | 95.87 |
| LLMs | **Qwen3-8B** | 96.85 | 99.01 | 95.40 | 97.04 |
| | **Mistral-7B** | **96.99** | 98.62 | **95.89** | 97.16 |

**TABLE 2 Class-wise performance of the fine-tuned models on the held-out test set**

| Group | Model | Class | Precision (%) | Recall (%) | F1-score (%) |
|---|---|---|---|---|---|
| **SLMs** | **Qwen3-1.7B** | No attack | 95.26 | **100.00** | **97.57** |
| | | Overshoot attack | **100.00** | 86.41 | 92.71 |
| | | Stopped attack | **100.00** | **100.00** | **100.00** |
| | | Turn-by-turn attack | **100.00** | **100.00** | **100.00** |
| | | Wrong-turn attack | **100.00** | **91.57** | **95.60** |
| | **Llama-3.2-1B** | No attack | 94.62 | **100.00** | 97.24 |
| | | Overshoot attack | **100.00** | **88.35** | **93.81** |
| | | Stopped attack | **100.00** | **100.00** | **100.00** |
| | | Turn-by-turn attack | 93.75 | **100.00** | 96.77 |
| | | Wrong-turn attack | **100.00** | 84.34 | 91.50 |
| **LLMs** | **Qwen3-8B** | No attack | 95.05 | **100.00** | 97.46 |
| | | Overshoot attack | **100.00** | 85.44 | 92.15 |
| | | Stopped attack | **100.00** | **100.00** | **100.00** |
| | | Turn-by-turn attack | **100.00** | **100.00** | **100.00** |
| | | Wrong-turn attack | **100.00** | **91.57** | **95.60** |
| | **Mistral-7B** | No attack | **95.67** | 99.53 | 97.56 |
| | | Overshoot attack | **100.00** | **88.35** | **93.81** |
| | | Stopped attack | **100.00** | **100.00** | **100.00** |
| | | Turn-by-turn attack | **100.00** | **100.00** | **100.00** |
| | | Wrong-turn attack | 97.44 | **91.57** | 94.41 |

The class-wise results in **Table 2** show that all models achieved perfect precision, recall, and F1-score for the stopped attack. Qwen3-1.7B, Qwen3-8B, and Mistral-7B also achieved perfect performance for the turn-by-turn attack, whereas Llama-3.2-1B obtained an F1-score of 96.77% because of its lower precision. For the no-attack class, Qwen3-1.7B achieved the highest F1-score of 97.57% and correctly identified all normal observations, indicating that the framework produced no false alarms. Although all models achieved 100% precision for the overshoot attack, their recall ranged from 85.44% to 88.35%, indicating that some overshoot samples were misclassified as other classes. Llama-3.2-1B and Mistral-7B achieved the highest overshoot F1-score of 93.81%. For the wrong-turn attack, Qwen3-1.7B and Qwen3-8B obtained the highest F1-score of 95.60%, with a recall of 91.57%. Overall, the results indicate that our structured narrative representation enables both SLMs and LLMs to distinguish normal vehicle behavior from multiple GNSS spoofing attack scenarios with high accuracy. More importantly, Qwen3-1.7B achieved the highest macro F1 Score despite having considerably fewer parameters than the evaluated LLMs. This finding suggests that a properly fine-tuned SLM can serve as an effective alternative to larger models for GNSS spoofing detection.

**Computational Efficiency and Resource Utilization**

Although the evaluated SLMs and LLMs achieved comparable detection performance, their computational requirements differ substantially. Therefore, inference latency, execution time, model size, GPU memory consumption, and training time are evaluated to examine the practical suitability of each model for GNSS spoofing detection. All models were evaluated using the same test samples and the same structured input and output-generation procedure. The experiments are conducted using NVIDIA A100 GPUs. The implementation is developed in Python using PyTorch 2.9.0+cu128, Hugging Face Transformers 5.14.1, Accelerate 1.14.0, and the Parameter-Efficient Fine-Tuning library for LoRA-based adaptation. The software environment also included Tokenizers 0.22.2, Safetensors 0.8.0, Hugging Face Hub 1.24.0, and SentencePiece 0.2.2.

*Inference Efficiency*

**Table 3** presents the inference-time performance of the evaluated models. Qwen3-1.7B achieved the lowest reported average inference latency of 122.69 ms per test sample and an average end-to-end latency of 123.65 ms. It required 85.64 s to generate labels and completed the entire test set in 86.79 s. Importantly, the average end-to-end latency of every evaluated model remained below 1 second. Because the data are sampled at 10 Hz and each non-overlapping observation window contains 10 samples, a new window becomes available every 1 second. Therefore, all evaluated models can process an observation window before the subsequent window is fully acquired, preventing a processing backlog under the evaluated conditions. Qwen3-1.7B provides the largest timing margin, completing the full inference pipeline in approximately 123.65 ms and leaving approximately 876.35 ms before the next observation window becomes available.

Qwen3-1.7B reduced the average inference latency by 22.29% compared with Qwen3-8B and by 81.62% compared with Mistral-7B. Equivalently, Mistral-7B required approximately 5.44 times longer than Qwen3-1.7B to generate a prediction for an individual observation window. The overall runtime of Qwen3-1.7B was also 22.09% lower than that of Qwen3-8B and 81.44% lower than that of Mistral-7B.

Llama-3.2-1B achieved the shortest model-loading time at 6.34 s, followed by Qwen3-1.7B at 8.21 s. In comparison, Mistral-7B and Qwen3-8B required 15.93 s and 24.06 s, respectively. The shorter loading times of the SLMs are advantageous when the detection model must be initialized frequently or deployed on computing platforms with limited resources.

**TABLE 3 Inference efficiency of the evaluated SLMs and LLMs on test samples**

| | Model | Average Inference Time (ms) | Minimum Time (ms) | Maximum Time (ms) | Average End-to-End Latency (ms) | Total Generation Time (s) | Overall Test Runtime (s) | Model Load Time (s) |
|---|---|---|---|---|---|---|---|---|
| **SLMs** | Qwen3-1.7B | **122.69** | 110.98 | **186.33** | **123.65** | **85.64** | **86.79** | 8.21 |
| | Llama-3.2-1B | 150.34 | **76.24** | 279.59 | 151.73 | 104.94 | 105.98 | **6.34** |
| **LLMs** | Qwen3-8B | 157.89 | 142.72 | 240.12 | 158.92 | 110.21 | 111.40 | 24.06 |

| | Mistral-7B | 667.69 | 360.56 | 727.95 | 669.22 | 466.05 | 467.61 | 15.93 |
|---|---|---|---|---|---|---|---|---|

*Computational Resource Requirements*

**Table 4** summarizes the model sizes, training times, and GPU memory requirements. As expected, the SLMs contained substantially fewer parameters and consumed considerably less GPU memory than the two LLMs. Llama-3.2-1B had the smallest model size, containing approximately 1.24 billion parameters. It required 2.32 GB of GPU memory after loading and reached a peak inference memory consumption of only 2.33 GB. Llama-3.2-1B achieved the lowest peak training GPU memory consumption at 3.09 GB, followed by Qwen3-1.7B at 3.30 GB. Despite not having the lowest training-memory requirement among the evaluated models, Qwen3-1.7B required 79.89% less peak training memory than Qwen3-8B (16.41GB) and 76.09% less than Mistral-7B (13.80 GB).

**TABLE 4 Model size, GPU memory consumption, and fine-tuning time**

| | Model | Total Parameters (Billion) | GPU Memory After Loading (GB) | Peak Inference GPU Memory (GB) | Training Time (hours) | Training Time (min) | Peak Training GPU Memory (GB) |
|---|---|---|---|---|---|---|---|
| **SLMs** | Qwen3-1.7B | 1.73 | 3.23 | 3.25 | 0.129 | 7.73 | 3.30 |
| | Llama-3.2-1B | **1.24** | **2.32** | **2.33** | **0.120** | **7.23** | **3.09** |
| **LLMs** | Qwen3-8B | 8.21 | 15.32 | 15.35 | 0.367 | 22.02 | 16.41 |
| | Mistral-7B | 7.26 | 13.56 | 13.59 | 1.095 | 65.71 | 13.80 |

Qwen3-8B required 15.35 GB of peak inference GPU memory, while Mistral-7B required 13.59 GB. Qwen3-1.7B therefore reduced peak inference memory consumption by 78.83% relative to Qwen3-8B and by 76.09% relative to Mistral-7B. The lower inference-memory requirements of the SLMs increase their suitability for deployment on embedded or resource-constrained computing platforms.

The SLMs also required substantially less fine-tuning time. Llama-3.2-1B completed fine-tuning in approximately 7.23 min, while Qwen3-1.7B required approximately 7.73 min. In contrast, Qwen3-8B required 22.02 min, and Mistral-7B required 65.71 min. Thus, Qwen3-1.7B reduced training time by 64.9% compared with Qwen3-8B and by 88.24% compared with Mistral-7B.

Qwen3-1.7B also achieved the lowest peak training GPU memory consumption at 2.30 GB. Its peak training memory was 85.99% lower than that of Qwen3-8B and 83.33% lower than that of Mistral-7B. These reductions demonstrate the practical benefit of parameter-efficient adaptation when comparatively small models are used for application-specific tasks, such as GNSS spoofing detection.

*Performance–Efficiency Trade-off*

The computational results must be considered together with the detection performance reported in Tables 1 and 2. Qwen3-1.7B achieved the highest macro F1-score of 97.18%, while Qwen3-8B and Mistral-7B achieved macro F1-scores of 97.04% and 97.16%, respectively. Therefore, increasing the number of parameters from 1.73 billion to approximately 7–8 billion did not produce a corresponding improvement in detection performance. Compared with Qwen3-8B, Qwen3-1.7B contained 78.93% fewer parameters, required 78.83% less peak inference memory, reduced average inference latency by 22.29%, and reduced training time by 64.9%, while achieving a slightly higher macro F1-score. Compared with Mistral-7B, Qwen3-1.7B contained 76.2% fewer parameters, required 76.09% less peak inference memory, reduced average inference latency by 81.62%, and reduced training time by 88.24%, while achieving a marginally higher macro F1-score.

**Figure 4** illustrates the relationship between detection performance and computational efficiency. Models positioned toward the upper-left region provide the preferred operating characteristics because they combine a high macro F1-score with low inference latency. The marker size represents peak inference GPU memory. Qwen3-1.7B occupies the most favorable region of the figure, achieving the highest macro F1-score, the lowest reported average inference latency, and substantially lower memory consumption than the evaluated LLMs. Overall, Qwen3-1.7B provided the most favorable detection-performance and inference-latency trade-off, achieving the highest observed macro F1-score and the lowest average inference-latency. Llama-3.2-1B required the least memory and training time but produced lower detection performance. Thus, Qwen3-1.7B was selected because it provided a stronger balance between predictive performance and inference responsiveness, while still requiring substantially fewer resources than the evaluated LLMs.

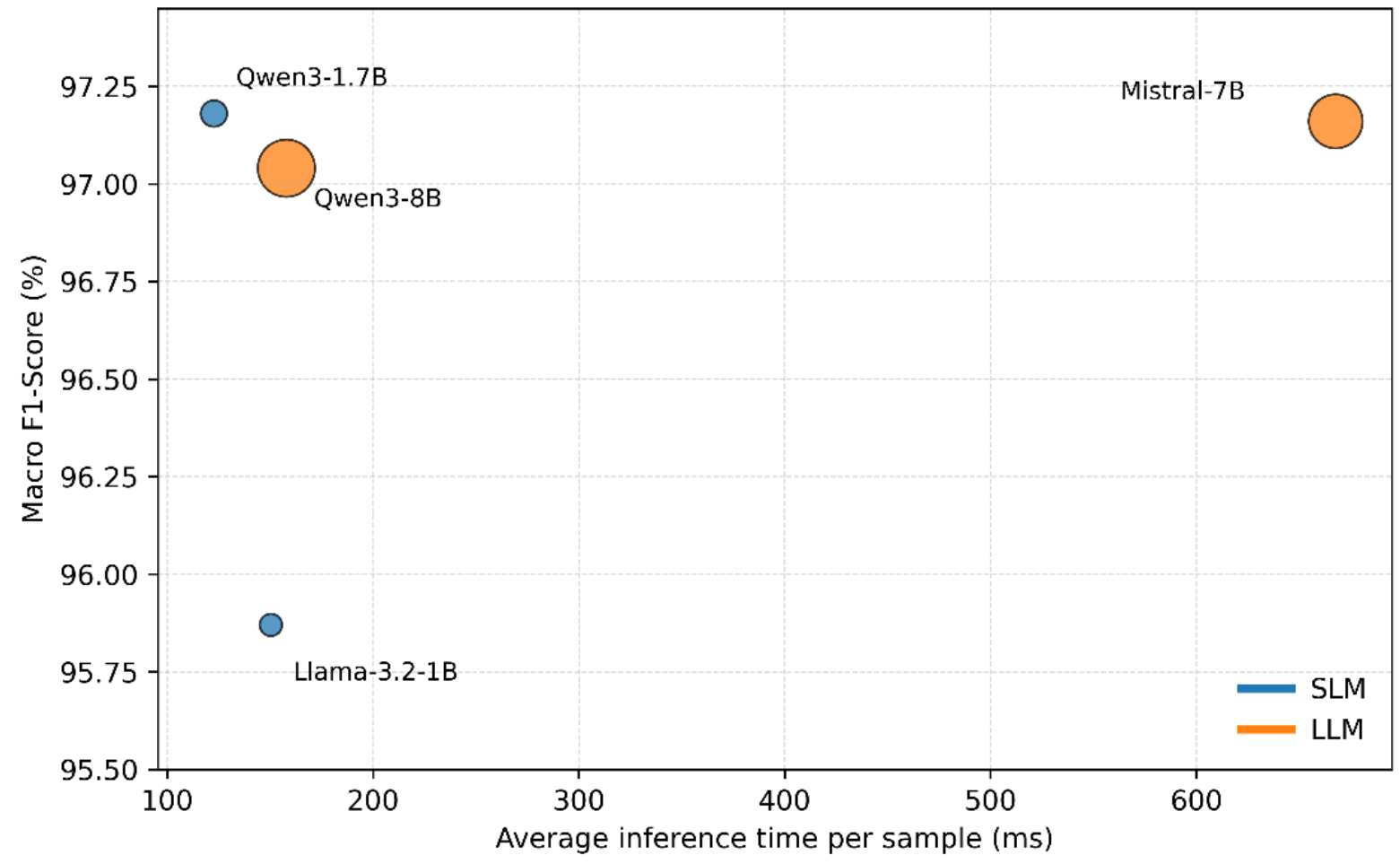

**Figure 4 Performance-efficiency trade-off among the evaluated SLMs and LLMs. The marker area is proportional to peak inference GPU memory**

**Evaluation on Unseen Field Data**

As shown in **Figure 4**, Qwen3-1.7B provides the most favorable performance–efficiency trade-off among the evaluated SLMs and LLMs. It achieves the highest macro F1-score while requiring the lowest average inference latency and substantially less GPU memory than the larger models. Therefore, Qwen3-1.7B is selected for an additional field-test evaluation to investigate whether a model fine-tuned using KITTI data collected in Germany could generalize to data collected in a geographically different region. The model is evaluated using field data collected in Clemson, South Carolina, United States. The class-wise and macro-average performance results are summarized in **Table 5**. On the field-test dataset, the model achieves an overall accuracy of 95.89%, along with a macro-average precision of 95.89%, recall of 96.07%, and F1-score of 95.89%. These results are comparable to those obtained on the original held-out test set, for which the model achieves an accuracy of 96.99% and macro F1-score of 97.18%. The relatively small reduction in performance indicates that the relationships learned between GNSS-derived and IMU-derived vehicle states are not limited to the geographical environment represented in the fine-tuning dataset.

**TABLE 5 Class-wise performance of Qwen3-1.7B on the unseen Clemson, U.S. field-test dataset**

| Class | Precision (%) | Recall (%) | F1-score (%) |
|---|---|---|---|
| **No attack** | 93.66 | 97.56 | 95.57 |
| **Overshoot attack** | 100.00 | 95.14 | 97.51 |
| **Stopped attack** | 95.35 | 95.35 | 95.35 |
| **Turn-by-turn attack** | 90.48 | 100.00 | 95.00 |
| **Wrong-turn attack** | 100.00 | 92.31 | 96.00 |
| **Macro Average** | 95.89 | 96.07 | 95.89 |

Overall, the field-test results provide evidence that the SLM-based detection framework can transfer the learned relationships between GNSS- and IMU-derived vehicle states to the evaluated geographically distinct dataset. The results suggest that the model captures sensor-consistency patterns associated with vehicle behavior; however, evaluations across additional geographic regions and driving conditions would be required to establish broader geographic generalizability.

## CONCLUSIONS

This study developed an SLM-based framework for detecting and classifying sophisticated GNSS spoofing attacks in autonomous vehicles by identifying inconsistencies between vehicle states independently derived from GNSS and IMU measurements. The framework presented in this paper converts motion, maneuver, speed, and heading information into structured semantic representations, enabling a compact language model to learn both numerical and semantic inconsistency patterns associated with five scenarios: no-attack, overshoot, stopped, turn-by-turn, and wrong-turn scenarios.

Experimental results demonstrate that SLMs can achieve spoofing detection performance similar to that of substantially larger language models while offering clear computational advantages. The best-performing SLM achieved an accuracy of 96.99%, precision of 99.05%, recall of 95.59%, and F1-score of 97.18%. Furthermore, the evaluated SLMs required lower inference latency and less GPU memory during fine-tuning and inference than the LLMs. This favorable balance between detection performance and computational efficiency represents a key advantage of SLMs over LLMs for practical vehicular applications. These findings suggest that increasing model size is not necessarily required to achieve effective spoofing detection when sensor measurements are transformed into a structured, task-specific representation.

The framework presented in the paper was further evaluated using field data collected in Clemson, South Carolina, United States, which were geographically independent of the KITTI data used for fine-tuning. The comparable performance achieved on this unseen dataset indicates that the framework primarily learns transferable relationships between multi-sensor vehicle behaviors rather than location-specific characteristics. Overall, the framework presented in this paper provides an accurate, computationally efficient, and geographically transferable solution for GNSS spoofing detection. Its low latency and memory requirements make it a promising candidate for near-real-time deployment on resource-constrained vehicular computing platforms.

## ACKNOWLEDGMENTS

This work is based upon the work supported by the National Center for Transportation Cybersecurity and Resiliency (TraCR) (a US Department of Transportation National University Transportation Center) headquartered at Clemson University, Clemson, South Carolina, USA. Any opinions, findings, conclusions, and recommendations expressed in this material are those of the author(s) and do not necessarily reflect the views of TraCR, and the US Government assumes no liability for the contents or use thereof.

ChatGPT was used only to help improve grammar. No information has been generated using any Large Language Model or Generative Artificial Intelligence. The large language models Qwen3-8B and Mistral-7B-Instruct-v0.3 and the small language models Qwen3-1.7B and Llama-3.2-1B-Instruct were fine-tuned and evaluated as part of the proposed GNSS spoofing detection framework, as described in the methodology of this paper.

## AUTHOR CONTRIBUTIONS

The authors confirm their contribution to the paper as follows: All authors reviewed the results and approved the final version of the manuscript.

## DECLARATION OF CONFLICTING INTERESTS

The authors declared no potential conflicts of interest.

## FUNDING

This work is based upon the work supported by the National Center for Transportation Cybersecurity and Resiliency (TraCR), under Grants: 69A3552344812, 69A3552348317.